\documentclass[%
reprint,
superscriptaddress,
amsmath,amssymb,
aps,
]{revtex4-1}
\usepackage[pdftex]{graphicx}
\usepackage{dcolumn}
\usepackage{bm}
\usepackage{float}
\usepackage{color}

\begin{document}

\title{Two floating camphor particles interacting through the lateral capillary force}

\author{Yuhei Hirose}
\affiliation{Department of Applied Physics, Graduate School of Science, Tokyo University of Science, 6-3-1 Niijuku, Katsushika-ku, Tokyo 125-8585, Japan}

\author{Yusuke Yasugahira}
\affiliation{Research Institute for Electronic Science, Hokkaido University, Sapporo 001-0020, Japan}

\author{Mamoru Okamoto}
\affiliation{Research Institute for Electronic Science, Hokkaido University, Sapporo 001-0020, Japan}

\author{Yuki Koyano}
\affiliation{Department of Physics, Tohoku University, 6-3, Aoba, Aramaki, Aoba-ku, Sendai 980-8578, Japan}

\author{Hiroyuki Kitahata}
\affiliation{Department of Physics, Chiba University, 1-33 Yayoi-cho, Inage-ku, Chiba 263-8522, Japan}

\author{Masaharu Nagayama}
\affiliation{Research Institute for Electronic Science, Hokkaido University, Sapporo 001-0020, Japan}

\author{Yutaka Sumino}
\email{ysumino@rs.tus.ac.jp}
\affiliation{
	Department of Applied Physics, Faculty of Science Division I, Tokyo University of Science, 6-3-1 Nijuku, Katsushika-ku, Tokyo, 125-8383, Japan}
\affiliation{
	Water Frontier Science \& Technology Research Center, I$^2$ Plus, and Division of Colloid and Interface Science, Research Institute for Science \& Technology, Tokyo University of Science, 6-3-1 Nijuku, Katsushika-ku, Tokyo, 125-8585, Japan
}

\date{\today}
\begin{abstract}
We consider a mathematical model for a two-particle system driven by the spatial gradient of a concentration field of chemicals with conservative attractive interactions in one dimension.
This setup corresponds to an experimental system with floating camphor particles at a water surface.
Repulsive interaction is introduced, as well as self-propelling force, through the concentration field of camphor molecules at the water surface.
Here we newly adopt the attractive lateral capillary force due to the deformation of the water surface.
The particles experience competing dissipative repulsion and conservative attraction.
We numerically investigated the mathematical model, and found six different modes of motion.
The theoretical approach revealed that some of such mode transitions can be understood in terms of bifurcation.
\end{abstract}

\maketitle

\section{Introduction}

Active matter is already a wide-spread concept, which treats a group of motile elements\cite{Howse,maass} as a novel type of matter\cite{Marchetti2013}. Active matter includes a group of biological objects, such as a school of fish, a flock of birds, and a colony of cells\cite{ViscekPhysRep}. Therefore, it attracts interest even from non-physicists.
Motility induced phase separation is one of the concepts found in active matter\cite{Cates2015}. The active Brownian particles\cite{Klimontovich1994, Schweitzer1998}, particles self-propelled with a finite speed under external noise, show phase separation\cite{PhysRevLett.108.235702, Stenhammar2013, Cates2015} only with local repulsive interaction. 
The system shows dynamic spatio-temporal patterns due to the competition of conservative local repulsion and dissipative driving force, which violates momentum conservation. It is notable that such competition is peculiar to active matter where far-from-equilibrium condition is imposed.

Dissipative effects can induce particle-particle interaction, as well as self-propulsion. Here, we focus our attention to the interaction through concentration fields. One of the examples is an interaction between cells\cite{Brown1974, Mato1975, Caterina1991} through diffused molecules. These cells release molecules to their environment, and each cell reacts to the concentration fields of the released molecules. The reaction includes the change of their self-propulsion, and additional release and/or consumption of the molecules. Finally, these cells spontaneously form a macroscopic bacterial colony with well-organized structures\cite{Budrene1991, Budrene1995,matusita68.1436,matusita65.2700}. 

Apart from biological systems, a camphor-water system is a well-studied self-propelled system where the particle can interact through a concentration field\cite{NakataLangmuir, NaktaPCCP,SelfOrganizedMotion}. Camphor is a surface active chemical with sublimability.
When a camphor particle is put at a water surface, camphor molecules are released from the particle, and reduce the surface tension of the water surface. Sublimation of camphor prevents the saturation of the water surface with the camphor molecules.
When a symmetrically shaped camphor particle, e.g.~a circular-shaped particle, is at rest, the surface tension balances around the particle.
Thus the rest state of a symmetric camphor particle can exist.
By perturbating the resting camphor particle, the concentration field of camphor molecules becomes asymmetric.
Such an asymmetric profile of the concentration field can drive the particle through the surface tension.
A positive feedback loop between the motion and the asymmetry in the concentration profile leads to continuous motion when the resistant force is small enough.
These processes are described with a simple mathematical model based on a reaction-diffusion equation\cite{Hayashima2001,Nagayama2004,koyano,koyano2,BonifacePhysRevE.99.062605,koyano_rotor}. The same type of model can be applied for other self-propelling chemical systems\cite{Bekki1990, Bekki1992, Nagai2005, Sumino2005b} driven by the imbalance of surface tension. The model can also be regarded as a model for a chemotactic motion, in which a particle moves in the direction of the gradient of chemical concentration, and thus it is not only a model for the camphor-water system, but also a generic model for the system in which the particles move depending on the concentration field around them\cite{Mikhailov}.

When multiple camphor particles are placed on a water surface, they interact through the concentration field of camphor molecules\cite{Kohira2001, Schulz, Grzybowski1,Grzybowski2,EI201810,ParmanandaPhysRevE,IkedaPhysRevE.99.062208, MOROHASHI2019104}.
The interaction is essentially repulsive, since they are driven in the direction with a lower concentration of camphor molecules. In this case, the repulsive interaction is effective in the range of the diffusion length $\ell_D$, determined by the diffusion coefficient $D$, and sublimation rate $\alpha$, as $\ell_D=(D/\alpha)^{1/2}$ \cite{koyano}. It should be noted that the concentration gradient of the surface active chemical component induces Marangoni convection\cite{Marangoni}, and it transports the chemical component itself. Due to this effect, the transport of the chemical component is enhanced. This effect can be included in the ``effective'' diffusion term under the assumption that the system is close to the steady state\cite{JCPMarangoni,quantitativeCamphor}, and $D$ in the present paper is the effective diffusion coefficient including the Marangoni effect. This mathematical model is commonly used to describe self-propulsion in the camphor-water system. Many characteristic collective behaviors of multiple camphor particles can be described with the same type of mathematical model including the effective diffusion coefficient\cite{Nishimori2017,Ikura, Nishi,suematsuJPSJ}.

\begin{figure}[bt!]
    \centering
    \includegraphics[scale=1]{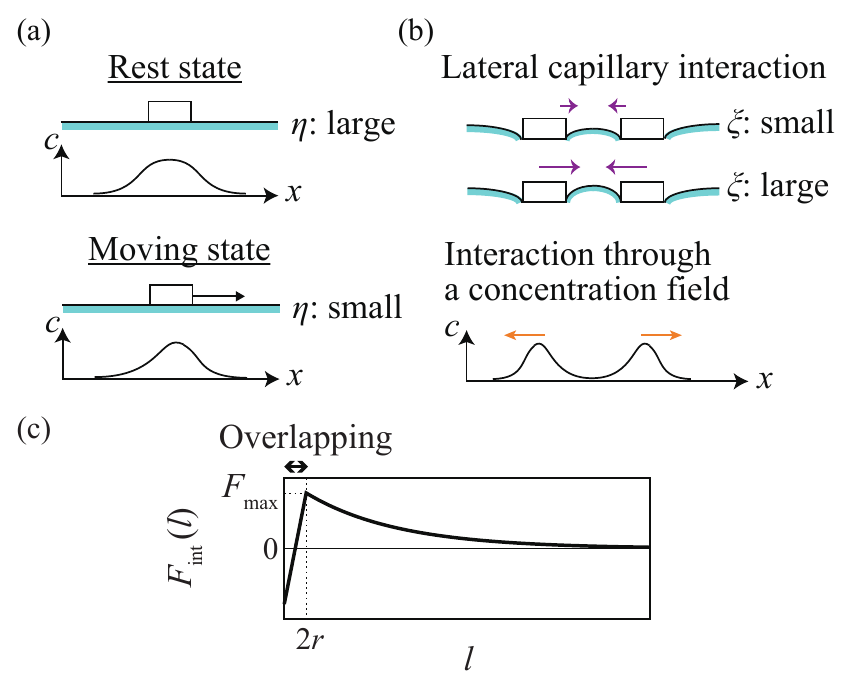}
    \caption{(a) Schematic illustration of a camphor particle and a concentration field. The particle stays static at large $\eta$, while it is driven and keeps moving at small $\eta$. (b) Schematic illustration of an interface deformed by two camphor particles and a concentration field. $\xi$ represents the amplitude of the attractive lateral capillary force. Repulsive interaction also works on both particles due to the concentration field. (c) Plot of the lateral capillary force $F_\textrm{int}$ described in Eq.~\eqref{Fint}. The value $F_\mathrm{max}$ is defined as $F_\mathrm{max} = \xi e^{-2qr}$.}
    \label{fig:my_label}
\end{figure}

Interestingly, any particles floating at a surface show conservative interaction due to the deformation of the surface; so-called lateral capillary force\cite{4, Kralchevsky2001}. The lateral capillary force appears to minimize the surface energy of water, and its direction is determined by the types of meniscus: convex and concave. The lateral capillary force is attractive (repulsive) for the same (different) type of meniscus. The lateral capillary force is effective in the range of the capillary length $q^{-1}$, determined by surface tension $\gamma$, density $\rho$, and gravitational acceleration $g$, as $q^{-1}=\left(\gamma/(\rho g) \right)^{1/2}$ \cite{deGennes}. Particles with similar physical characteristics, such as wettability and density, have the same type of meniscus and, hence, the interaction is attractive.

Overall, the conservative attractive and dissipative repulsive forces can compete in the camphor-water systems. As mentioned, such competition can be relevant to active matter system, but overlooked up to now. 
In this study, we consider the modes of motion for two camphor particles in a one-dimensional system.
To clarify the competition between the attractive and repulsive forces, we adopt the theoretical approach, where the balance of forces is easy to change.
The present paper starts from the introduction of the mathematical model for the motion of camphor particles. Then, the modes of motion depending on the balance between attractive and repulsive forces are numerically investigated.
Finally, some aspects of numerical results are explained by theoretical analysis.

\section{Mathematical Model}

We introduce a one-dimensional system with two camphor particles floating at a water surface based on the previous work \cite{Hayashima2001,Nagayama2004,koyano,koyano2}. The particles interact with each other through the camphor concentration field and the lateral capillary force. In this model, the time development of the camphor particle positions $x_i(t)$ ($i = 1, 2$) and the camphor surface concentration field $c(x,t)$ are considered.

The dynamics for the $i$-th camphor particle $x_i(t)$ ($x_1 < x_2$) is described as below: 
\begin{align}
    \begin{split}
    m\dfrac{d^2 x_1}{dt^2}=-\eta \dfrac{dx_1}{dt}-w \Gamma \left[c(x_1+r,t)-c(x_1-r,t)\right]\\
    +F_{\rm int}(l),
    \end{split}\\
    \begin{split}
    m\dfrac{d^2 x_2}{dt^2}=-\eta \dfrac{dx_2}{dt}-w \Gamma \left[c(x_2+r,t)-c(x_2-r,t)\right]\\
    -F_{\rm int}(l).
    \end{split}
\end{align}
where $l$ denotes the distance between the two particles
\begin{align}
l = x_2 - x_1. \label{def_l}
\end{align}
Here, $m$, $\eta$, and $r$ are the mass, friction coefficient, and radius of a camphor particle\cite{quantitativeCamphor}. In order to compromise dimensionality of the equation, here we introduce $w$ as a length scale. This $w$ corresponds to the width, the dimension perpendicular to the considered direction $x$ and the direction of the gravitational acceleration in a three-dimensional system. 
In the above equations, we assume the linear relation between the surface tension $\gamma$ and the camphor surface concentration $c$ as
\begin{align}
    \gamma = \gamma_0 - \Gamma c,
\end{align}
where $\gamma_0$ is the surface tension of pure water and $\Gamma$ is a positive constant.

$F_{\rm int}(l)$ reflects the lateral capillary force\cite{4, Kralchevsky2001} between two camphor particles:
\begin{align}
F_{\rm int}(l)=
\begin{cases}
    2w\gamma_0  e^{-q (l-r)} \sin^2 \psi, & l > 2r,\\
    -2w\gamma_0 e^{-q r} \dfrac{(2-\epsilon) r-l}{\epsilon r} \sin^2 \psi, & l \leq 2r.
\end{cases}
\label{Fint}
\end{align}
Here, $\psi$ is the contact angle of the water surface around the camphor disk, and $q$ is the inverse of the capillary length. We assume the surface tension modulation by the camphor concentration, $\Gamma$, is sufficiently small, and we do not consider the dependence of the camphor surface concentration in the calculation of the lateral capillary force.
A short-range excluding volume effect of the particle is included when $l \leq 2r$. We define that $F_{\rm int}$ is continuous at $l = 2r$. $\epsilon$ is a small parameter, and the $1 / \epsilon$ controls the strength of the repulsive force.
We define the characteristic intensity of the lateral capillary force as $\xi = 2w\gamma_0 e^{qr} \sin^2 \psi$.

Next, we consider the dynamics of the camphor surface concentration $c(x,t)$:
\begin{align}
    \frac{\partial c}{\partial t}=D \dfrac{\partial^2 c}{\partial x^2}-\alpha c+\sum_{i=1}^{2}S(x,x_i),
\end{align}
where $D$ and $\alpha$ denote the diffusion coefficient of camphor molecules at the water surface and sublimation rate. $S(x,x_i)$ represents a supply rate of camphor molecules from the $i$th particle located at $x_i$:
\begin{align}
        S(x,x_i)=
    \begin{cases}
    \dfrac{\beta}{2wr}, &\left| x-x_i\right| \leq r,\\
    0, &\left|x-x_i\right| > r,\\
    \end{cases}
\end{align}
where $\beta$ is a supply rate of camphor particles per unit time.
Since $S$ denotes the release per unit time and area, $\beta$ is divided by the area of a camphor particle, $2wr$.

In the numerical calculation and theoretical analysis, we adopt the dimensionless form of the model. The dimensionless variables and coefficients are given as:
\begin{align}
    &\tilde{x}=\sqrt{\dfrac{\alpha}{D}}x,& &\tilde{t}=\alpha t,&
    &\tilde{c}=\dfrac{D}{\beta}c,& \nonumber \\
    &\tilde{S}=\dfrac{D}{\alpha \beta}S,&
    &\tilde{\eta}=\dfrac{\eta}{m\alpha},& 
    &\tilde{\Gamma}=\dfrac{\beta\Gamma}{m\alpha^2 D},& \nonumber \\
    &\tilde{q}=\sqrt{\dfrac{D}{\alpha}}q,&
    &\tilde{r}=\sqrt{\dfrac{\alpha}{D}}r,&
    &\tilde{w}=\sqrt{\dfrac{\alpha}{D}}w,& \nonumber\\
    &\tilde{l}=\sqrt{\dfrac{\alpha}{D}}l,&
    &\tilde{\xi} = \dfrac{\xi}{m\alpha \sqrt{\alpha D}}.& \nonumber
\end{align}
The tildes ( $\tilde{}$ ) are omitted hereafter for simplicity.
Then, the following dimensionless equations are obtained:
\begin{align}
\begin{split}
    \dfrac{d^2 x_1}{dt^2}=-\eta \dfrac{dx_1}{dt}-w\Gamma \left[c(x_1+r,t)-c(x_1-r,t)\right]\\
    +F_\textrm{int}(l), \label{eq2a}
\end{split}\\
\begin{split}
    \dfrac{d^2 x_2}{dt^2}=-\eta \dfrac{dx_2}{dt}-w\Gamma \left[c(x_2+r,t)-c(x_2-r,t)\right]\\
    -F_\textrm{int}(l), \label{eq2b}
\end{split}
\end{align}
\begin{align}
    F_{\rm int}(l)=
\begin{cases}
    \xi e^{-q l}, & l> 2r,\\
    \\
    -\xi e^{-2q r} \dfrac{(2-\epsilon) r-l}{\epsilon r}, &l \leq 2r,
\end{cases} \label{force}
\end{align}
\begin{align}
    \dfrac{\partial c}{\partial t}=\dfrac{\partial^2 c}{\partial x^2}-c+\sum_{i=1}^{2}S(x,x_i), \label{eq1}
\end{align}
\begin{align}
    S(x,x_i)=
    \begin{cases}
    \dfrac{1}{2wr},&\left|x-x_i\right| \leq r,\\
    0, &\left|x-x_i\right| > r.\\
    \end{cases} \label{source}
\end{align}

\section{Numerical Results}

\begin{figure}[tb!]
    \centering
    \includegraphics[scale=1.0]{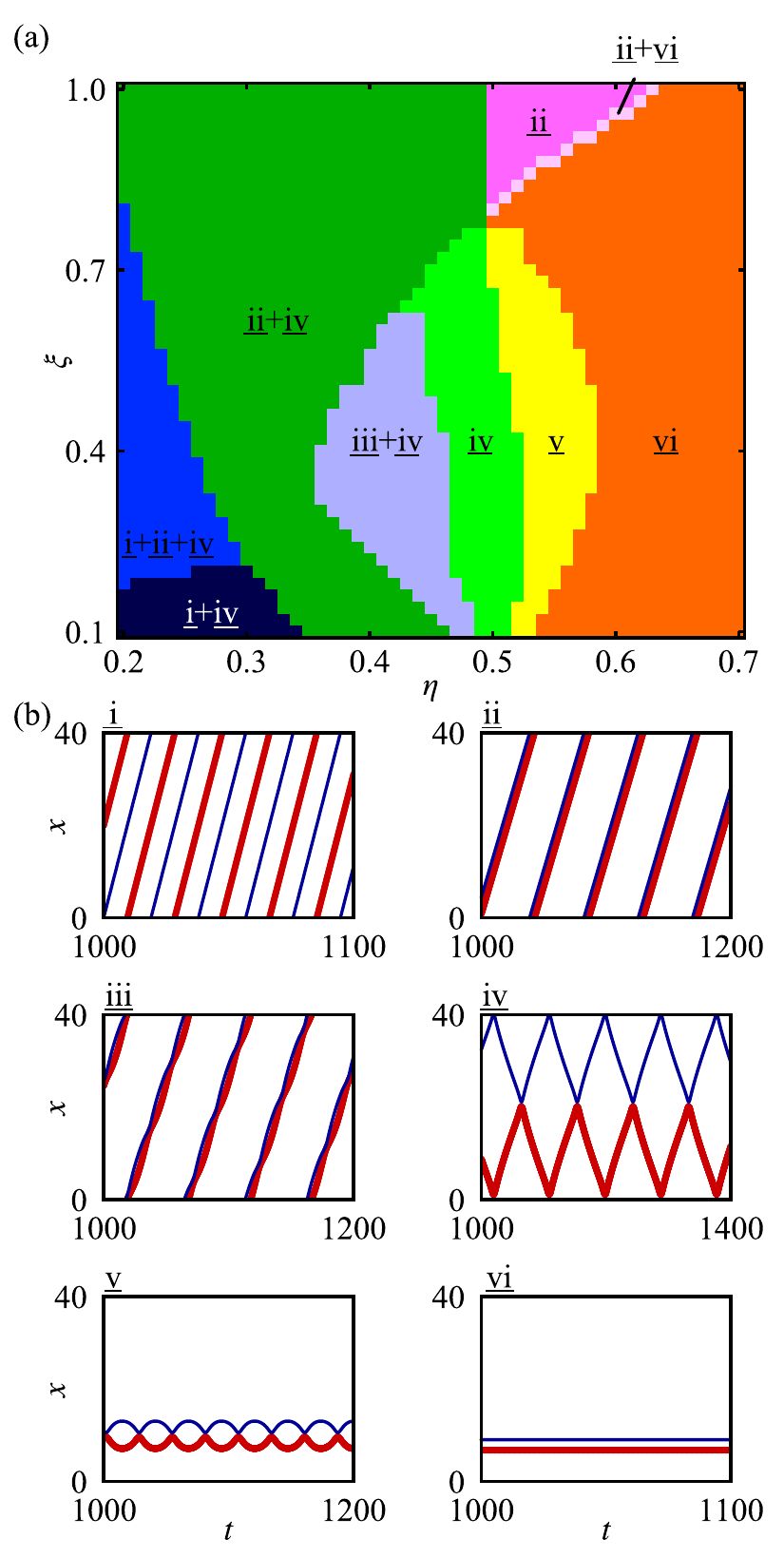}
    \caption{(a) Phase diagram for stable modes numerically obtained. We observed six characteristic modes. Each mode was labeled as \underline{i} isolated translational motion, \underline{ii} clustered translational motion, \underline{iii} inchworm motion, \underline{iv} head-on-collision, \underline{v} standing oscillation, and \underline{vi} standing cluster. Several stable modes can be observed with the same parameters (denoted by the sign ``+''). Controlled parameters were $\eta$ and $\xi$. The other parameters were fixed as $q=0.4$, $w=1.0$, $r=0.3$, $\Gamma=5.0$, $\epsilon=0.001$, and $L=40.0$. (b) Typical spatio-temporal plots for modes in (a). Red and blue lines represent the positions of the two particles. The values of $\eta$ and $\xi$ for the spatio-temporal plots are $(\eta,\xi)$ = \underline{i}:$(0.30,0.20)$, \underline{ii}:$(0.40,0.20)$, \underline{iii}:$(0.40,0.44)$, \underline{iv}:$(0.50,0.20)$, \underline{v}:$(0.55,0.40)$, and \underline{vi}:$(0.60,0.40)$.}
    \label{phase_diagram_num}
\end{figure}

The numerical calculation was performed to overview the behavior of a two-particle system. We adopted a one-dimensional system with a periodic boundary condition to investigate the long-term behavior without the effect of the system boundary. The system size was set as $L$. Some equations should be modified to reflect the periodic boundary condition. We introduce a function $\textrm{mod}(z)$ for a simple representation:
\begin{align}
    z = k L + \textrm{mod}(z), \quad k \in \mathbb{Z}, 
\end{align}
and
\begin{align}
    0 \leq \textrm{mod}(z) < L.
\end{align}
Equations
\eqref{eq2a} and \eqref{eq2b} are modified as
\begin{align}
    \dfrac{d^2 x_1}{dt^2}=&-\eta \dfrac{dx_1}{dt} \nonumber \\
    &-w \Gamma \left[c(\textrm{mod}(x_1+r),t)-c(\textrm{mod}(x_1-r),t)\right] \nonumber \\
    &+F_\textrm{int}(\textrm{mod}(x_2-x_1)) - F_\textrm{int}(\textrm{mod}(x_1-x_2)), 
\label{eq2_na}\\
    \dfrac{d^2 x_2}{dt^2}=&-\eta \dfrac{dx_2}{dt} \nonumber \\
    &-w \Gamma \left[c(\textrm{mod}(x_2+r),t)-c(\textrm{mod}(x_2-r),t)\right] \nonumber \\
    &+F_\textrm{int}(\textrm{mod}(x_1-x_2)) - F_\textrm{int}(\textrm{mod}(x_2-x_1)),
\label{eq2_nb}
\end{align}
and Eq.~\eqref{source} is modified as
\begin{align}
        S(x,x_i)=
    \begin{cases}
    \dfrac{1}{2wr}, & \min(\textrm{mod}(x-x_i), \textrm{mod}(x_i-x))\leq r,\\
    0, & \min(\textrm{mod}(x-x_i), \textrm{mod}(x_i-x)) > r .\\
    \end{cases} \label{source_n}
\end{align}
In addition, $x_i(t)$ ($i = 1,2$) is modified by adding or subtracting $L$ to be in the range of $0 \leq x_i(t) < L$ reflecting the periodic boundary condition, i.e., $x_i(t)$ is replaced with $\textrm{mod}(x_i(t))$ if $x_i(t) < 0$ or $x_i(t) \geq L$. It should be noted that $x_1$ can be greater than $x_2$, different from the original model, reflecting a periodic boundary condition. 
In the numerical calculation, the definition of $l$ is also modified to be 
\begin{align}
l = \min (\textrm{mod}(x_1-x_2), \textrm{mod}(x_2-x_1)) \label{def_lmod}.
\end{align}

The Crank-Nicolson scheme was used for Eqs.~\eqref{eq1} and \eqref{source_n}, while the Euler method was used for Eqs.~\eqref{eq2_na} and \eqref{eq2_nb}.
Time step and spatial mesh size were set to be $10^{-3}$ and $10^{-2}$, respectively. Fixed parameters were $q=0.4$, $w=1.0$, $r=0.3$, $\Gamma=5.0$, $\epsilon=0.001$, and $L=40.0$. In this study, we fixed $q<1$ so that the competition between conservative and dissipative interactions could be observed. 
We fixed $\Gamma$ and varied $\xi$ and $\eta$ as parameters, since $\Gamma$ represents the amplitude of the force and the ratio between $\eta$ and $\Gamma$ is important.  The values of the other parameters $w$, $\epsilon$, $L$, and $r$ do not affect the bifurcation structure; we set $\epsilon \ll r < 1 \ll L$, where the diffusion length is set to be unity.

We numerically investigated the dependence of the modes on the friction coefficient $\eta$ and the intensity of the lateral capillary force $\xi$ to observe the stable modes of camphor particle motion. We obtained the largest parameter region for each mode of motion in the following two manners: The one was to scan $\xi$ with an interval of $0.02$ in both (increasing and decreasing) directions while $\eta$ was fixed, and the other was to scan $\eta$ in both directions with an interval of $0.01$ while $\xi$ was fixed.
The initial values of $x_1$, $x_2$, $v_1$, and $v_2$ before scanning were set to be $x_1 = 0$, $x_2 = 5.0$ or $20.0$, $v_1 = v_2 = 0$, and $c(x) \equiv 0$. When the parameter values were changed during the scanning, we did not change the values of $x_1$, $x_2$, $v_1$, and $v_2$, but we added a spatially-incoherent noise to the concentration field $c(x)$ in order to make easier to escape from the unstable steady state. After the change in the parameters, we ran simulation long enough to be converged into characteristic modes.

The obtained phase diagram is shown in Fig.~ \ref{phase_diagram_num}.
There are characteristic modes of camphor particles motion, which are labeled by \underline{i} isolated translational motion, \underline{ii} clustered translational motion, \underline{iii} inchworm motion, \underline{iv} head-on-collision, \underline{v} standing oscillation, and \underline{vi} standing cluster.
The modes \underline{i} and \underline{iv} are characteristic to a finite-sized system and should not be present in an infinite system.

The modes \underline{i} and \underline{ii} are the translational motions of two camphor particles in the same direction.
The distance between the particles is $L/2$ and is smaller than $L/2$ for modes \underline{i} isolated translational motion and \underline{ii} clustered translational motion, respectively.
The mode \underline{iii} is the combination of translation and oscillation; so-called inchworm motion.
The modes \underline{iv} and \underline{v} are the oscillatory motion.
In the case of \underline{iv} head-on-collision, the amplitude of the oscillation is close to $L/2$, whereas the amplitude is smaller than $L/2$ in the case of \underline{v} standing oscillation.
Thus, the particles repeat head-on collisions twice in a period in the case of mode \underline{iv}. 
The mode \underline{vi} is the stationary state. The particles are localized such that the distance between particles is smaller than $L/2$.
There are bistable regions where several modes can be realized with different initial conditions as shown in Fig.~\ref{phase_diagram_num}(a).

\begin{figure}[bt]
    \centering
    \includegraphics[scale=1.0]{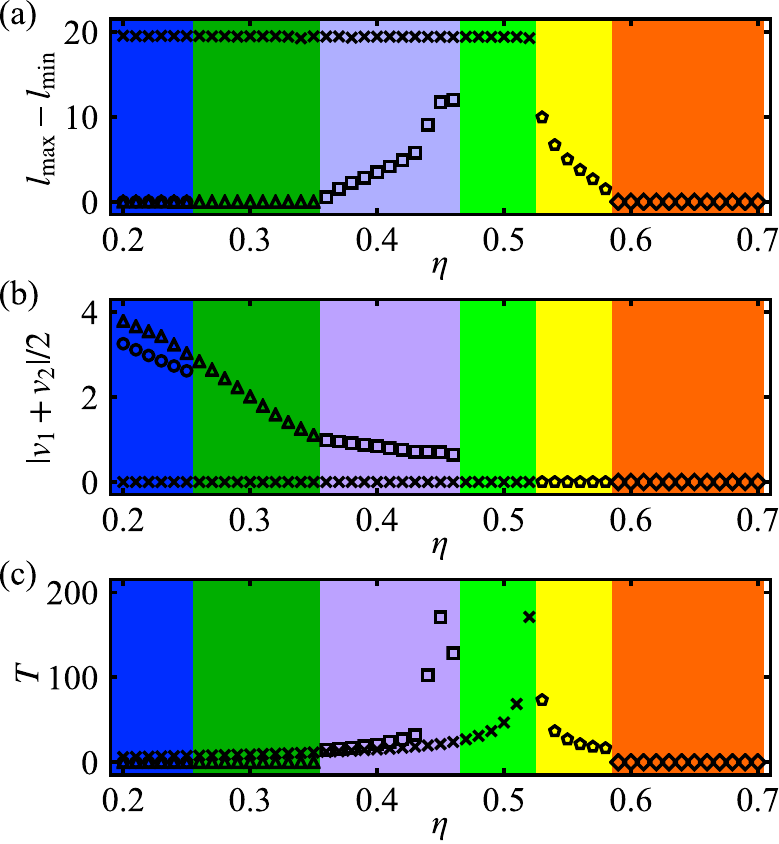}
    \caption{Order parameters depending on $\eta$ at $\xi=0.4$. (a) Amplitude of oscillation, $l_\textrm{max} - l_\textrm{min}$. (b) Average speed
    $\left|v_1 + v_2\right|/2$.  (c) Period of oscillation, $T$. Circles, triangles, squares, crosses, pentagons, and diamonds represent the values of \underline{i}, \underline{ii}, \underline{iii}, \underline{iv}, \underline{v}, and \underline{vi}, respectively.}
    \label{order4}
\end{figure}

\begin{figure}[bt]
    \centering
    \includegraphics[scale=1.0]{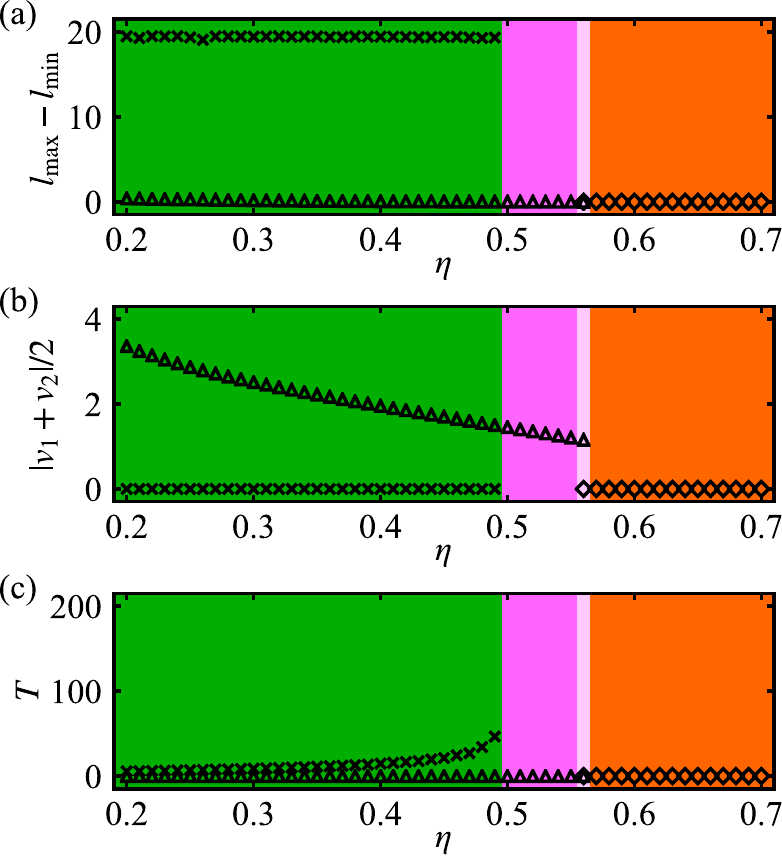}
    \caption{Order parameters depending on $\eta$ at $\xi=0.9$. (a) Amplitude of oscillation, $l_\mathrm{max} - l_\textrm{min}$. (b) Average speed $\left|v_1 + v_2\right|/2$.  (c) Period of oscillation, $T$. Triangles, crosses and diamonds represent the values of \underline{ii}, \underline{iv}, and \underline{vi}, respectively.}
    \label{fig:my_label2}
\end{figure}

We used three order parameters to classify the modes of motion: the difference between the distance of two particles $l_{\mathrm{max}}-l_{\mathrm{min}}$, the average speed $|v_1 + v_2|/2$, and the period $T$. $l_{\mathrm{max}}$ and $l_{\mathrm{min}}$ are the maximum and the minimum values of $l$ in appropriate time span after the sufficiently long annealing time. We denote $T=0$ for the modes without oscillation. 
The dependence of the order parameters on the friction coefficient $\eta$ is shown in Figs. \ref{order4} and \ref{fig:my_label2}, where $\xi$ = 0.4 and 0.9, respectively.

For $\xi=0.4$ (Fig.~\ref{order4}), three different modes can be stable at $\eta \leq 0.25$. Circles indicate \underline{i} isolated translational motion, where $l_\textrm{max} - l_\textrm{min}=0$ and average speed $\left|v_1 + v_2\right|/2$ is finite. These order parameters indicate that the two particles travel with a fixed distance at a finite speed. We confirmed the distance is fixed as 20, which corresponds to the half size of the system size $L/2$. Triangles, indicating \underline{ii} clustered translational motion, also show a similar behavior, while the average speed is slightly larger than the one for circles (\underline{i}). We also confirmed the distance is kept as smaller than $L/2$.
Crosses (\underline{iv} head-on-collision) show $l_\textrm{max} - l_\textrm{min} \simeq L/2 = 20$ and $\left|v_1 + v_2\right|/2=0$. Here, the particles show the oscillatory motion whose amplitude is approximately $L/2$.

\begin{figure}[bt]
    \centering
    \includegraphics[scale=1.0]{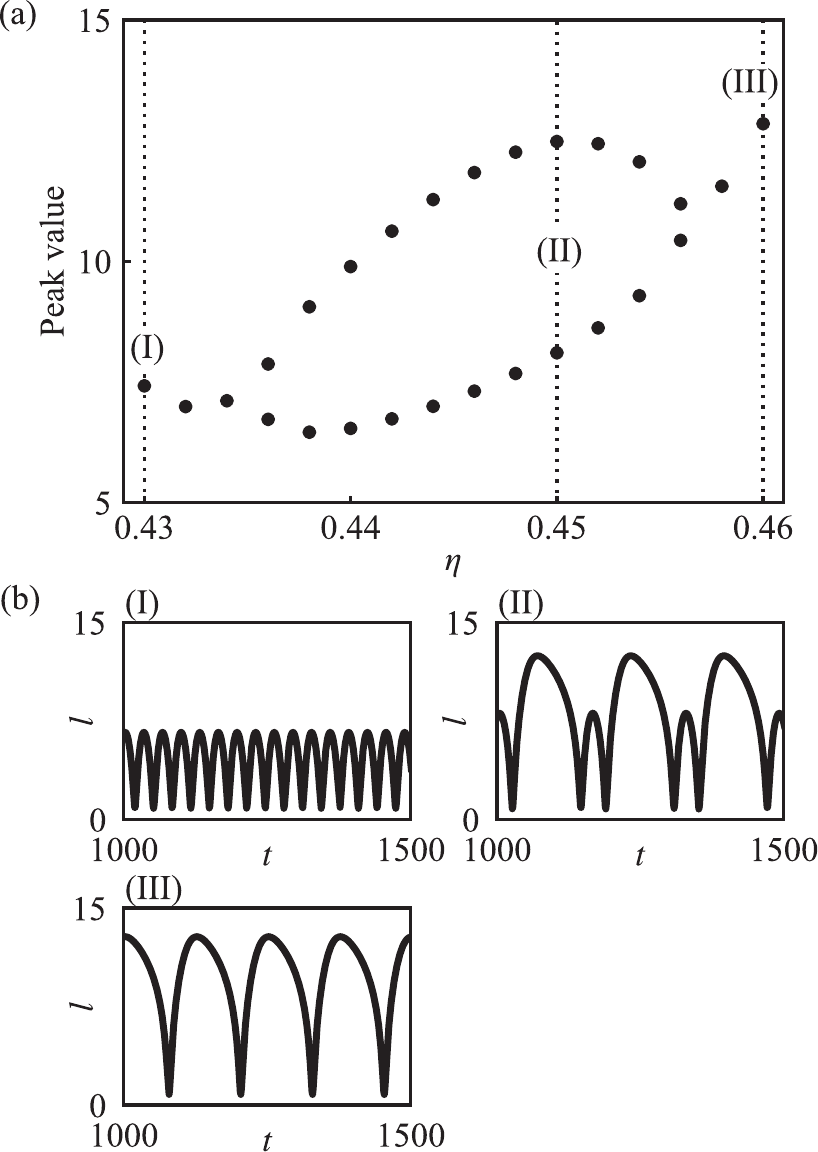}
    \caption{(a) Plots of peak values (=Local maximum distance(s) between two particles) depending on $\eta$. The parameters are the same as Fig.~\ref{order4}. 
    (b) Time series of the distance between two particles, $l$. $\eta=$ (I) 0.43, (II) 0.45, (III) 0.46. Figure \ref{fig:my_label4}a was obtained by the peak value(s) of each time series of the distance.}
    \label{fig:my_label4}
\end{figure}

For $0.26 \leq \eta \leq 0.35$, triangles (\underline{ii}) and crosses (\underline{iv}) can be observed.
By increasing $\eta$ further, squares (\underline{iii} inchworm motion) appear, instead of triangles (\underline{ii}); for $0.36 \leq \eta \leq 0.46$. These squares (\underline{iii}) are characterized by finite amplitude $l_\textrm{max} - l_\textrm{min}$ together with a finite average speed $\left|v_1 + v_2\right|/2$.
When $0.47 \leq \eta$, squares (\underline{iii}) disappear. Close to this transition line, we observed a long oscillation period $T$, which indicates the appearance of complex oscillation discussed later (Fig.~\ref{fig:my_label4}).

For $0.47 \leq \eta \leq 0.52$, only crosses (\underline{iv}) can be observed. Crosses (\underline{iv}) disappear for $0.53 \leq \eta$. Instead, pentagons (\underline{v} standing oscillation), or diamonds (\underline{vi} standing cluster) appear in the case for $0.53 \leq \eta \leq 0.58$ and $0.59 \leq \eta$, respectively. Pentagons (\underline{v}) are characterized by average speed $\left|v_1 + v_2\right|/2=0$, and a finite oscillation amplitude $l_\textrm{max} - l_\textrm{min}$. Here, $l_\textrm{max} - l_\textrm{min} < L/2$, different from crosses (\underline{iv}). Diamonds (\underline{vi}) are characterized by $\left|v_1 + v_2\right|/2 =0$ and $l_\textrm{max} - l_\textrm{min}=0$, while the distance between the particles was confirmed to be smaller than $L/2$, indicating the particles are clustered.   

For $\xi=0.9$ (Fig.~\ref{fig:my_label2}), the behavior of the system is qualitatively similar to the one shown in Fig.~\ref{order4}. Triangles (\underline{ii}) and crosses (\underline{iv}) were observed for $\eta \leq 0.49$. Only triangles (\underline{ii}) appear for $0.5 \leq \eta \leq 0.55$. With the coexistence at $\eta = 0.56$, there are only diamonds  (\underline{vi}) in $\eta \leq 0.57$.

In the case of squares (\underline{iii}), a period-doubling behavior was observed (Fig.~\ref{fig:my_label4}). The data shown correspond to $\xi=0.4$; the same parameters with Fig.~\ref{order4}. At $0.434 < \eta < 0.436$, period-doubling bifurcation is suggested to occur. For $0.436 \leq \eta \leq 0.446$, the oscillation has two peaks in the amplitude during a period. Similarly, another period-doubling bifurcation is suggested to occur at $0.456 < \eta < 0.458$.
The period increases monotonically till $\eta= 0.47$ as $\eta$ is increased. The combination of these two factors is a cause of the long oscillation period that appeared close to $\eta= 0.47$.

\section{Analytical Results}

We performed the theoretical analysis to understand the stable modes appearing in Fig.~\ref{phase_diagram_num} from the viewpoint of dynamical systems.

We should recall that the original model with infinite system size is described by the following equations:
\begin{align}
	\dfrac{\partial c}{\partial t}= \frac{\partial^2 c}{\partial x^2}-c+\sum_{i=1}^{2}S(x,x_i), \label{c_theoretical_analysis}
\end{align}
\begin{align}
	S(x,x_i)=
	\begin{cases}
	\dfrac{1}{2wr}, & |x-x_i|\leq r,\\
	0, & |x-x_i|> r,\\
	\end{cases}
\end{align}
\begin{align}
	\dfrac{d^2 x_1}{dt^2}&=-\eta \dfrac{d x_1}{dt} - w\Gamma[c(x_1+r,t)-c(x_1-r,t)]+\xi e^{-ql}, \label{d2x1dt2}\\
	\dfrac{d^2 x_2}{dt^2}&=-\eta \dfrac{d x_2}{dt} - w\Gamma[c(x_2+r,t)-c(x_2-r,t)]-\xi e^{-ql}, \label{d2x2dt2}
\end{align}
\begin{align}
l = x_2 - x_1. \label{def_l_th}
\end{align}
Here we should also recall $x_1 < x_2$.

First, we assume that the relaxation of the concentration field is much faster than the acceleration and deceleration of the particles.
We construct a concentration field around a single camphor particle moving with a constant speed $v_c$ in the moving frame $X=x-v_c t$. Then, the equation for a concentration field,
\begin{align}
	\frac{\partial c}{\partial t} = \frac{\partial^2 c}{\partial x^2} - c + S(x,v_c t), 
\end{align} leads an ordinary differential equation for $C(X, v_c) = c(X + v_c t,t)$ as\cite{Shimokawa}
\begin{align}
	-v_c \frac{d C}{d X} = \frac{d^2 C}{d X^2} - C + S(X,0).
\end{align}
The actual expression of the concentration field is obtained as follows:
\begin{align}
	C(X,v_c)=
	\begin{cases}
	-\dfrac{1}{wr}\dfrac{\kappa_-}{\kappa_+-\kappa_-}\sinh(\kappa_+r)e^{\kappa_+X}, & X<-r,\\
	\dfrac{1}{2wr} + \dfrac{1}{2wr}\dfrac{\kappa_-}{\kappa_+-\kappa_-}e^{\kappa_+(X-r)} \\
	\qquad -\dfrac{1}{2wr}\dfrac{\kappa_+}{\kappa_+-\kappa_-}e^{\kappa_-(X+r)}, & \left|X\right| \leq r,\\
	-\dfrac{1}{wr}\dfrac{\kappa_+}{\kappa_+-\kappa_-} \sinh(\kappa_-r)e^{\kappa_-X}, & X>r.\\
	\end{cases}
	\label{concentration_field_co-moving}
\end{align}
Here, we define $\kappa_\pm=\dfrac{-v_c\pm \sqrt{{v_c}^2+4}}{2}$.
By substituting Eq.~\eqref{concentration_field_co-moving} with Eqs.~\eqref{d2x1dt2} and \eqref{d2x2dt2}, and differentiating the both sides of Eq.~\eqref{def_l_th}, we have the dynamical system with three variables for the two-camphor-particle system as follows:
\begin{align}
	\dfrac{dv_1}{dt}=-\eta v_1 - w\Gamma[&C(r,v_1)-C(-r,v_1)+C(-l+r,v_2) \nonumber \\
	&-C(-l-r,v_2)]
	+\xi e^{-ql},
\label{dv1dt}
\end{align}
\begin{align}
	\dfrac{dv_2}{dt}=-\eta v_2 - w\Gamma[&C(r,v_2)-C(-r,v_2)+C(l+r,v_1) \nonumber \\
	&-C(l-r,v_1)]
	-\xi e^{-ql},
\label{dv2dt}
\end{align}
\begin{align}
	\dfrac{dl}{dt}=v_2-v_1, \label{dldt}
\end{align}
where $v_1$ and $v_2$ denote the velocities of the two particles, which are regarded as constant for the characteristic time scale of the change in concentration field.
Here we define the right-hand sides of Eqs.~\eqref{dv1dt}, \eqref{dv2dt}, and \eqref{dldt} as $dv_1/dt=F_1(v_1,v_2,l)$, $dv_2/dt=F_2(v_1,v_2,l)$, and $dl/dt=F_3(v_1,v_2,l)$.

First, we perform the linear stability analysis of the solution corresponding to the standing cluster. By setting $v_1 = v_2 = 0$ and
 solving 
$dv_1/dt = dv_2/dt = dl/dt = 0$, i.e., $F_1(0,0,l) = F_2(0,0,l) = F_3(0,0,l) = 0$, we have a stationary solution corresponding to the standing cluster as:
\begin{align}
    (v_1,v_2,l)=&\left (0,0,\dfrac{1}{1-q} \ln \left(\dfrac{\Gamma \sinh^2 r}{r\xi} \right)\right) \nonumber \\
    \equiv&(0,0,l_0). \label{stationalstanding}
\end{align}
Then Eqs.~\eqref{dv1dt}, \eqref{dv2dt}, and \eqref{dldt} are linearized around the stationary solution in Eq.~\eqref{stationalstanding} as
\begin{align}
\frac{d}{dt}
{\renewcommand\arraystretch{2}
	\begin{bmatrix}
	\delta v_1 \\
	\delta v_2 \\
	\delta l 
	\end{bmatrix}
}
=
{\renewcommand\arraystretch{2}
	\begin{bmatrix}
	\dfrac{\partial F_1}{\partial v_1} & \dfrac{\partial F_1}{\partial v_2} &\dfrac{\partial F_1}{\partial l} \\
	\dfrac{\partial F_2}{\partial v_1} & \dfrac{\partial F_2}{\partial v_2} &\dfrac{\partial F_2}{\partial l} \\
	\dfrac{\partial F_3}{\partial v_1} & \dfrac{\partial F_3}{\partial v_2} &\dfrac{\partial F_3}{\partial l} 
	\end{bmatrix}_{(v_1,v_2,l)=(0,0,l_0)}
}
{\renewcommand\arraystretch{2}
	\begin{bmatrix}
	\delta v_1 \\
	\delta v_2 \\
	\delta l 
	\end{bmatrix}
}.
\label{liniarized_equation}
\end{align}
The eigenvalues of the matrix appeared in Eq.~\eqref{liniarized_equation}, $\lambda$, are obtained as
\begin{align}
    \lambda = \sigma_1(\eta,\xi), \; \sigma_2(\eta,\xi) \pm i \omega_2(\eta,\xi),
\end{align}
where 
\begin{align}
    \sigma_1 = -\eta + \dfrac{\Gamma}{2r}
    &\left[e^{-r} (\sinh r - r e^{-r}) \right. \nonumber \\
     &\, \left.+ e^{-l_0} (r \sinh 2r - (1 + l_0) \sinh^2 r) \right],
\end{align}
\begin{align}
    \sigma_2 = -\frac{\eta}{2} + \dfrac{\Gamma}{4r}
    & \left[e^{-r} (\sinh r - r e^{-r}) \right. \nonumber \\
    &\, \left. - e^{-l_0} (r \sinh 2r - (1 + l_0) \sinh^2 r) \right],
\end{align}
\begin{align}
    \omega_2 = \sqrt{\omega_0^2 - {\sigma_2}^2}.
\end{align}
Here we define the frequency of oscillation $\omega_0$ at the Hopf bifurcation point ($\sigma_2 = 0$) as
\begin{align}
    \begin{split}
    \omega_0 = \sqrt{2 (1-q) \xi e^{-l_0 q}}.
    \end{split}
\end{align}

\begin{figure}[t]
    \centering
    \includegraphics{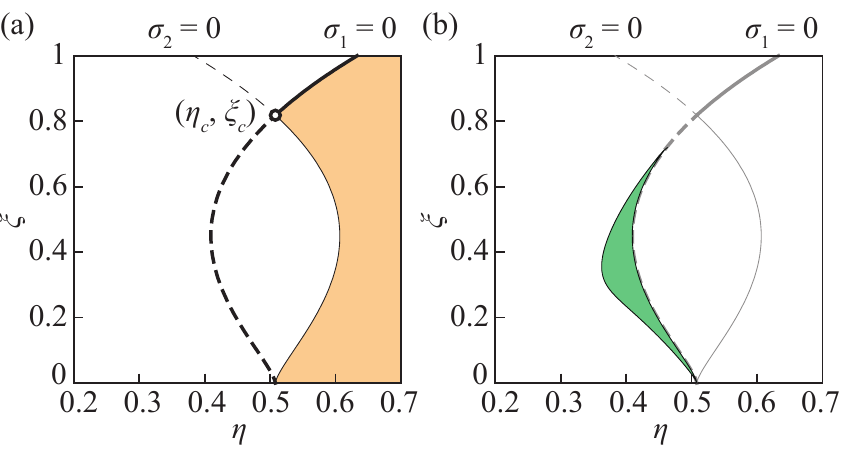}
    \caption{Phase diagram obtained from the theoretical analysis. (a) Phase diagram obtained from the linear stability analysis of the standing cluster (\underline{vi}). The standing cluster state (Eq.~\eqref{stationalstanding}) is linearly stable in the orange-colored parameter region, where both $\sigma_1$ and $\sigma_2$ are negative. The crossing point of the two curves $\sigma_1 = 0$ and $\sigma_2 = 0$ is set as $(\eta_c, \xi_c)$. For $\xi > \xi_c$, the standing cluster is destabilized through a pitchfork bifurcation at the thick corresponding to $\sigma_1=0$. For $0< \xi < \xi_c$, it is destabilized through a Hopf bifurcation at the thin line corresponding to $\sigma_2=0$. (b) Phase diagram obtained from the linear stability analysis of the clustered translational state (\underline{ii}). The green-colored region shows the region where the mode (\underline{ii}) exists but it is destabilized. The left-side curve corresponds to a Hopf bifurcation from clustered translational motion (\underline{ii}) to inchworm motion (\underline{iii}), whereas the right-side curve corresponds to the boundary of the existence of the clustered translational motion (\underline{ii}).
        }
    \label{phase_diagram_theoretical}
\end{figure}

By changing the values of $\xi$ and $\eta$, the sign of the real part of eigenvalue $\lambda$ changes.
When the signs of $\sigma_1$ and $\sigma_2$ are negative, the standing cluster (\underline{vi}) is linearly stable.
When the sign of $\sigma_1$ changes to positive, a pitchfork bifurcation occurs and a solution corresponding to the clustered translational motion can emerge.
On the other hand, when the sign of $\sigma_2$ changes to positive, a Hopf bifurcation occurs and a solution corresponding to the standing oscillation can emerge.
Figure \ref{phase_diagram_theoretical}(a) shows the phase diagram obtained by the linear stability analysis.
The curves $\sigma_1=0$ and $\sigma_2=0$ intersect at $(\eta_c, \xi_c)$ as shown in Fig. \ref{phase_diagram_theoretical}(a). It suggests that the standing cluster is destabilized through a pitchfork bifurcation for $\xi > \xi_c \simeq 0.818$, while it is destabilized through Hopf bifurcation for $\xi < \xi_c$.
Comparing Fig.~\ref{phase_diagram_theoretical}(a) with Fig.~\ref{phase_diagram_num}, it is said that the mode transition from the standing cluster (\underline{vi}) to the clustered translational motion (\underline{ii}) and standing oscillation (\underline{v}) is well reproduced.

Next, we also investigate the linear stability of the solution corresponding to the clustered translational motion. In the same way as in the case of the standing cluster, we set $v_1 = v_2 = v > 0$ and calculate the stationary solution $(v_1, v_2, l) = (v, v, l_1)$. 
Unfortunately, $l_1$ cannot be explicitly described as an analytic function of $v$, and thus we obtain a set of $(v,l_1)$ numerically. It should be noted that one set of $(v, l_1)$ is obtained for each set of $(\eta, \xi)$ when $\sigma_1 > 0$, and the values of $(v, l_1)$ are continuously connected to $(v, l_0)$ on a curve with $\sigma_1 = 0$, which means the pitchfork bifurcation at $\sigma_1 = 0$ is supercritical. Obviously, $(v_1, v_2, l) = (-v, -v, l_1)$ is also another solution corresponding to the clustered translational motion. Hereafter, we only focus on the solutions which appear through the supercritical pitchfork bifurcation at $\sigma_1 = 0$, though there may be other solutions.

In order to investigate the linear stability of the solution corresponding to the clustered translational motion, the eigenvalues of the matrix
\begin{align}
    B = {\renewcommand\arraystretch{2}
	\begin{bmatrix}
	\dfrac{\partial F_1}{\partial v_1} & \dfrac{\partial F_1}{\partial v_2} &\dfrac{\partial F_1}{\partial l} \\
	\dfrac{\partial F_2}{\partial v_1} & \dfrac{\partial F_2}{\partial v_2} &\dfrac{\partial F_2}{\partial l} \\
	\dfrac{\partial F_3}{\partial v_1} & \dfrac{\partial F_3}{\partial v_2} &\dfrac{\partial F_3}{\partial l} 
	\end{bmatrix}_{(v_1,v_2,l)=(v,v,l_1)}
}
\end{align}
are calculated numerically, and evaluated whether they are real or complex and the signs of their real parts. One of the three eigenvalues always has negative real value, while the others are complex conjugates and the signs change depending on the parameters. The results of the linear stability analysis is shown in Fig.~\ref{phase_diagram_theoretical}(b). In the case that the standing cluster solution is firstly destabilized through pitchfork bifurcation, i.e., $\xi > \xi_c$, the sign of the complex eigenvalues is negative, and a stable translational cluster motion occurs. In contrast, in the case that the standing cluster solution is firstly destabilized through Hopf bifurcation, i.e., $\xi < \xi_c$, the sign of the real part of the complex eigenvalues is negative for smaller $\eta$, while it is positive for greater $\eta$, and Hopf bifurcation occurs between them. This Hopf bifurcation corresponds to the transition from the clustered translational motion (\underline{ii}) for smaller $\eta$ to the inchworm motion (\underline{iii}) for greater $\eta$, which is also observed in numerical results.

It is noted that there are qualitative differences between phase diagrams obtained numerically and theoretically. This discrepancy should reflect that the numerical calculation is performed in a finite-sized system with a periodic boundary condition.

\section{Discussion and Summary}
In the present paper, we considered a mathematical model of camphor particles floating at a water surface as a system where conservative and dissipative driving force competes. 
A floating camphor particle experiences the force directing lower surface concentration of camphor molecules. A single camphor particle at a water surface shows self-propulsion by the spontaneous symmetry breaking. In addition, the concentration fields result in effective repulsion between particles. In this study, we newly adopted the lateral capillary force that act as a conservative attractive force between camphor particles.

Our numerical results revealed that two particles in one-diemsional system show six different modes, some of which are bistable with the same parameter for different initial conditions. Notably, we newly found inchworm like motion where the distance between particles oscillates while the center of mass of these particles shows a translational motion.

The natural extension of the present study is the Maranogoni-driven multiple-particle system with conservative attractive interactions in two dimension. Indeed, experimental study shows dynamic clustering of Marangoni-driven agents floating at a solution surface~\cite{Schulz, Grzybowski1,Grzybowski2,ShimpeiTanaka}. The extension of our model into multiple particles in two dimension is left for the future study. 

A concentration field is often used to transmit signals between actively moving elements. Bacteria release signaling molecules, which can be detected by bacteria themselves. They can move in response to the gradient of the signaling molecule; so-called chemotaxis. Our model, describing floating camphor particles, can be recognized as the model for such chemotactic behavior with conservative long-range attraction. We believe that a group of the diffusion-driven agents whose motion is stabilized due to spontaneous symmetry breaking, like the present model, should be recognized as a novel class of active matter.

\section*{ACKNOWLEDGMENTS}
The authors acknowledge J.~Gorecki for his helpful discussion.
This work was supported by JSPS KAKENHI Grant Numbers JP19J00365 JP16H03949, JP16K13866, JP16H06478, JP19H05403
and the Cooperative Research Program of ``Network Joint Research Center for Materials and Devices'' (Nos.~20181023, 20181048, 20183003, 20194006, and 20191030).
This work was also supported by JSPS and PAN under the Japan-Poland Research Cooperative Program ``Spatio-temporal patterns of elements driven by self-generated, geometrically constrained flows''.

\end{document}